# How many infections of COVID-19 there will be in the "Diamond Princess"- Predicted by a virus transmission model based on the simulation of crowd flow

Zhiming Fang[1,#,*], Zhongyi Huang[1,#], Xiaolian Li[2,#,*], Jun Zhang[3], Wei Lv[4], Lei Zhuang[5], Xingpeng Xu[1], Nan Huang[1]

1 Business School, University of Shanghai for Science and Technology, Shanghai 200093, China

2 College of Ocean Science and Engineering, Shanghai Maritime University, Shanghai 201306, China

3 State Key Laboratory of Fire Science, University of Science and Technology of China, Hefei 230026, China

4 Research Center for Crisis & Hazard Management, Wuhan University of Technology, Wuhan 430070, China

5 Shanghai Rules and Research Institute, China Classification Society, Shanghai 200135, China

# These authors contributed equally to this work
* Corresponding Author: Zhmfang2015@163.com; lixl@shmtu.edu.cn

**Abstract:**

**Objectives:** Simulate the transmission process of COVID-19 in a cruise ship, and then to judge how many infections there will be in the 3711 people in the "Diamond Princess" and analyze measures that could have prevented mass transmission.

**Methods:** Based on the crowd flow model, the virus transmission rule between pedestrians is established, to simulate the spread of the virus caused by the close contact during pedestrians' daily activities on the cruise ship.

**Measurements and main results:** Three types of simulation scenarios are designed, the **Basic scenario** focus on the process of virus transmission caused by a virus carrier and the effect of the personal protective measure against the virus. The condition that the original virus carriers had disembarked halfway and more and more people strengthen self-protection are considered in the **Self-protection scenario**, which would comparatively accord with the actual situation of "Diamond princess" cruise. **Control scenario** are set to simulate the effect of taking recommended or mandatory measures on virus transmission

**Conclusions:** There are 850~1009 persons (with large probability) who have been infected with COVID-19 during the voyage of "Diamond Princess". The crowd infection percentage would be controlled effectively if the recommended or mandatory measures can be taken immediately during the alert phase of COVID-19 outbreaks.

**Keywords:** COVID-19, Diamond Princess, cruise ship, virus transmission model, virus infection rate

# 1 Introduction

A novel coronavirus (COVID-2019) broke out in mainland of China, Wuhan city, since Dec. 2019 and fast spread to the other areas of China and the other countries along with the Chinese Spring Festival holiday. To mitigate the spread of the virus, Chinese government and its Chinese people took a quick response: the Wuhan city announced its sealing off at first, then, the Chinese government announced to extend the holiday, after that, Chinese local governments announced to postpone the work returning successively and strongly suggested people to stay at home as far as possible. Numerous domestic airports and train stations have adopted temperature screening measures to detect individuals with fever (Wu et al., 2020).

Despite of a series of prevention and control measures, according to CNN, until 21ST Feb. confirmed cases of the novel coronavirus worldwide still reaches a number more than 70,000, bringing its total death toll to more than 2,000 (Helen and Adam, 2020). Compared to the SARS-CoV, emerged in 2002, which caused more than 8,000 infections and 800 deaths, and MERS-CoV, emerged in 2012, which caused 2494 infections and 858 deaths worldwide, COVID-2019 has higher potential for sustained community transmission (Wu et al., 2020), which increases the difficulty to prevent and control. This fact is unceasingly verified by Chinese new infections and unusually large case clusters, and further zoomed into the eyes of the whole world at the "Diamond Princess" cruise ship.

The "Diamond Princess" cruise ship is a truly luxurious cruise ship worldwide, whose tonnage is large to 115,875. It has 18 decks with 1,337 guest cabins, which gives it a capability to accommodate 2,670 guests and 1,100 crews (https://www.princess.com/ships-and-experience/ships/di-diamond-princess/). The ship departed from Yokohama (YOK), Japan on 20th Jan. and carried one guest from Hong Kong who has been reported as the first infection on 1st Feb. On 3rd Feb, the ship returned to YOK and is put into quarantine, and on 9th Feb, those whose virus detection result is negative disembarked from the ship. According to China News, until to 20th Feb, confirmed cases of the novel coronavirus at the ship reaches a number of 634 (http://www.chinanews.com/gj/2020/02-21/9099900.shtml), that is, after the first infection disembarked from the ship, 633 cases were confirmed as infections, leading the infection ratio higher to 17% which is far more than the average infection ratio of other communities. The explanation for reasons of the high ratio is numerous, such as the relatively closed space, the high dependency of the central air-conditioning system, and the ineffective quarantine (https://www.maritime-executive.com/article/nih-official-diamond-princess-quarantine-failed). These reasons are taken reasonable, that is, in these days, relatively closed space, numerous public spaces, and ineffective control measures give the virus suitable conditions to make them keep spreading like invaders.

From the scientific aspect, there are a lot of members at the ship and the interaction among the members is relatively simple for the few contact with external people, plus with the intact, detailed record of the infection cases, which makes the ship an ideal source to study and simulate the virus spreading in communities. However, the main infection process is still unclear until now, even, a fact that it happened before the quarantine or after it is currently subject to some debates. So, investigating the virus spreading process by using an infectious disease transmission model would be an effective reference for this up in the air thing.

Referring to the theory in Ref. (Zhang et al., 2011), the present models of infectious disease transmission can be divided into three categories according to the, which are single group model (treat the population as a whole), the composite group model (divide population into child groups according to the spatial heterogeneity) and individual models (consider attributes and behaviors of individuals). The SIR model (Kermack and McKendrick, 1991) proposed in 1927 is a classical single group model, in which the population was treated as a whole and partial differential equations was adopted to describe the relationship between healthy population, infected population, recovering population and dead population. Besides, the concept of average contact number was introduced and they pointed out that the average contact number was the key factor affecting the spread of infectious diseases. Another single group model proposed by Hoppensteadt et al. (1975) considered the effect of age heterogeneity on the transmission. Random factors were then introduced by some studies (Black and McKane, 2010; Nåsell, 2002), and these models were more effective than deterministic models in exploring the characteristics of the early stage of a infectious diseases and the recurring infectious disease. Composite group models divide the population into different subgroups according to clearly defined social units. With the hypothesis, this kind of models are able to investigate the transmission of infectious diseases caused by personnel movement in different areas with the help of the graph theory. Cross et al. (2007) established a $11 \times 11$ grid model. A SIR model was adopted in the each grid. Another composite group model was proposed by Colizza and Vespignani (2008) to obtain threshold conditions for epidemic infection, in which a SIR model was used to describe flows within subgroups, and a network was used to describe links between subgroups. Watts et al. (2005) treated the population as a hierarchical structure and established a composite group model to discuss the impact of hierarchy on the transmission.

In individual models, however, the basic modeling unit is individual rather than a group of people. Personal attributes and behavior rules are assigned to each individual, and then contact network will be established between individuals. Kleczkowsi et al. (1999) introduced the cellular automata to the simulation of the epidemic process of infectious diseases, and discussed the impact of personnel flow and medical treatment on epidemic transmission. Based on the survey data of personal contact in large cities, Eubank et al. (Eubank et al., 2004) developed EpiSimS. With the software they studied the spread and control of smallpox in cities. Milne et al. (2008) established a virtual network model based on the population, household characteristics and location information of an Australian town. The model described the contact and infection process of personnel, and studied the intervention effect of prevention and control measures on influenza.

In this paper, we simulate, discuss and predict the viral transmission process on the "Diamond Princess" based on a crowd flow model, with which behaviors, contacts and infections of people can be described at individual level. So straightly speaking, it belongs to the individual model introduced above. In Section 2 we introduce the foundational assumptions and rules of the model, and then the results of the simulation are discussed in Section 4. At last, the summary is presented in Section 4.

## 2 Model

### 2.1 The simulation of crowd flow

The method to simulate the flow process of the people on the cruise ship is as follows:

- Several action forms for pedestrians are defined, including going to the restaurant for meals (three meals a day), going to the leisure area (irregularly), and returning to the room to rest (irregularly during the day, at night).

- Each pedestrian is given a series of "time – action form", for example, "7:05 - going to the restaurant", "9:23 – going to the leisure area", … , "21:40 - returning to the room".

- When a certain time in the "time – action form" is reached, the corresponding action form will be implemented, take the pedestrian with "7:05 - going to the restaurant" as an example, he will start to move towards a restaurant at the 7:05.

- The pedestrians' movement rules during each action form are derived from the movement rules of the discrete evacuation model (Fang et al., 2010). The difference is that the destination of pedestrians in the evacuation model is the exit, yet the destination here is the restaurant, leisure area, room, etc.

**2.2 The virus transmission model based on the simulation of crowd flow**

Based on the crowd flow model, the virus transmission rule between pedestrians is established, to simulate the spread of the virus caused by the close contact during pedestrians' daily activities in the cruise ship. The virus transmission model is defined as follows:

- First, $d_{vir}$ is used to represent the viral propagation distance, $R_{per-vir}$ is the probability of virus infection of each pedestrian.
- When the distance between the virus carrier and surrounding pedestrians is less than $d_{vir}$, the probability of virus infection of each surrounding pedestrian $R_{per-vir}^{round}$ is determined by equation (1).

$$R_{per-vir}^{round} = \max(R_{per-vir}^{round}, f(i_{day})f(\Delta t)(1-S_{per}^{round})(1-S_{per}^{center})R_{pers-vir}^{center}) \quad (1)$$

Where, $S_{per}^{center}$, $S_{per}^{round}$ are the effectiveness of protective measures adopted by the virus carrier and each surrounding pedestrian respectively. $R_{pers-vir}^{center}$ is the probability of virus infection of the virus carrier. $f(i_{day})$ is a function of the number of days of the virus incubation ($i_{day}$) of the virus carrier. We assumed that during the virus incubation period ($I_{per}$), the virus transmission efficiency increases linearly, as shown in equation (2).

$$f(i_{day}) = \min(i_{day}/I_{per}, 1) \quad (2)$$

$f(\Delta t)$ is a function of the contact time with virus carriers ($\Delta t$). We assumed that the virus transmission efficiency increases linearly as $\Delta t$ increases, as shown in equation (3).

$$f(\Delta t) = \min(\Delta t/t_{con}, 1) \quad (3)$$

Where, $t_{con}$ is a confirm time within which close contact will definitely cause the virus to spread.

- There could be multiple original virus carriers in the model, whose $R_{per-vir}$ are 1.

- Considering that some people may be immune to the virus, a maximum value of 50% is assumed for the $R_{per-vir}$ of the original normal pedestrians, which means that 50 percent of people are immune to the virus.
- Referring to the characteristics of COVID-19, $d_{vir}$ is assumed as 1m, $I_{per}$ in equation (2) is assumed as 7, and $t_{con}$ in equation (3) is assumed as 2 minutes.
- $N_{peron}$ represents the total number of pedestrians in the ship. Each pedestrian may or may not use personal protective measure, and then the protection ratio $P_{protect}$ is defined which means that $P_{protect} \cdot N_{peron}$ pedestrians adopt personal protective measure. Moreover, we assumed that the effect of personal protective measure is equivalent to the KN90 mask, thus $S_{per}$ in equation (1) is 0.9 if pedestrian adopt personal protective measure, and 0 if not.
- $N_{day}$ represents the sailing days of the ship, which is also the number of days simulated in the model. When the simulation is finished, the final virus infection ratio in all pedestrians ($R_{crowd-vir}$) is calculated by equation (4).

$$R_{crowd-vir} = {\sum R_{per-vir}} / {N_{peron}} \tag{4}$$

## 3 Simulation Results

### 3.1 Object of the simulation

"Diamond Princess" is the top luxury cruise, which has 18 decks in total. Due to the unknown information of structure of cruise, it is difficult to simulate the COVID-19 infected situation on "Diamond Princess" cruise or on the similar cruise. Through careful investigation, it is found that the structure drawing of cruise ship (shown in Fig.1) that is named "President VI" of Vista Cruise is public on the Internet. For simplicity, the virus transmission model was established by using the structure graphing of "President VI", then the final virus infection ratio ($R_{crowd-vir}$) of "President VI" could be obtained and used as an analogy to analyze the infected situation of "Diamond Princess".

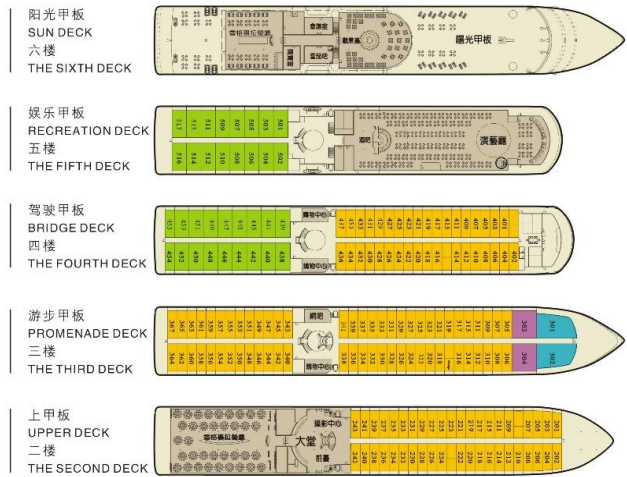

**Fig.1.** This figure comes from the Internet site
(*http://www.vst88.com/Youlun/ChuanZhiCJ.aspx?id=97cd8b90-77fd-4deb-bda8-bb1a681ddeae*)

As shown in Fig.2, the model for the cruise ship in Fig. 1 is built, in which there are 157 rooms, 2 pedestrians per room, and thus $N_{peron}$=354. Furthermore, the links of 2 animations are given in the appendix section, which present the dynamic process of crowd flow and virus transmission on the cruise ship.

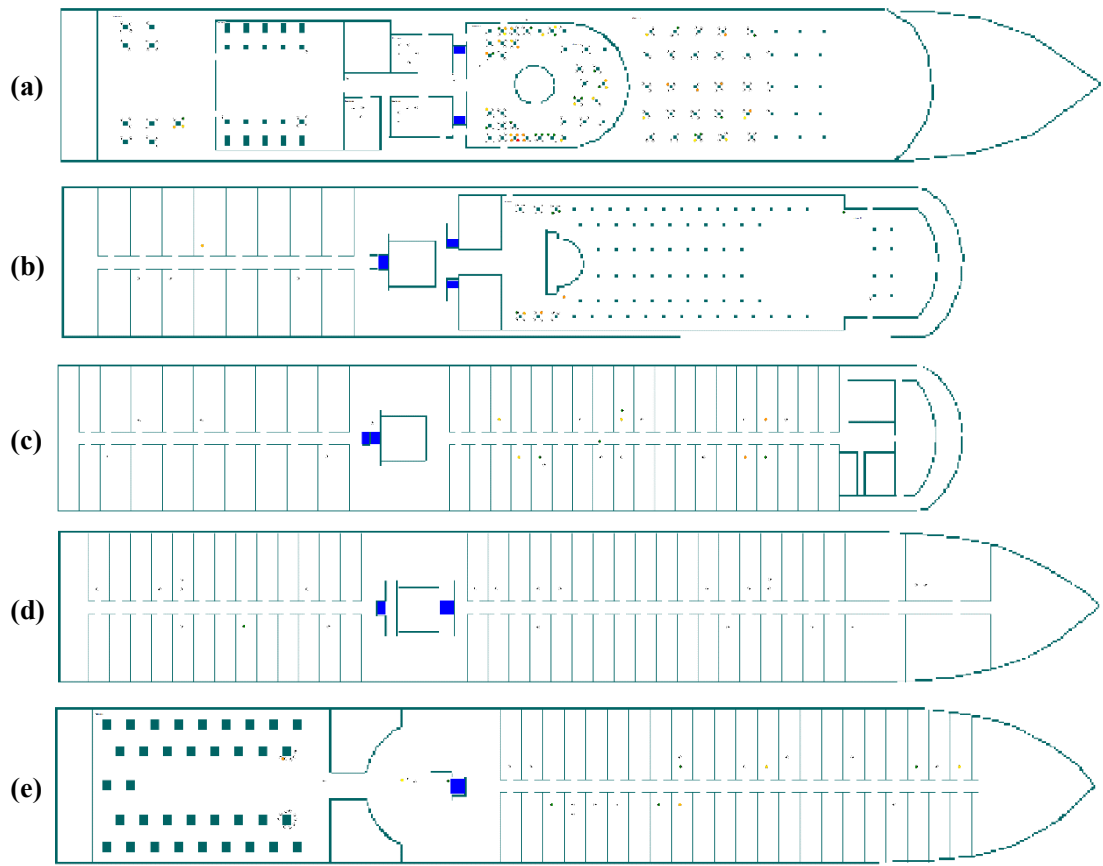

**Fig. 2.** Models of the cruise ship shown in Fig. 1. The dynamic process could be found in the links in appendix section.

## 3.2 Basic Scenarios

It is necessary to review the voyage of "Diamond Princess" cruise and key points of COVID-19 outbreaks before simulating.

*Jan. 20*: "Diamond Princess" cruise departed from Yokohama, Japan.

*Jan. 23*: The government took some measures to limit the inflow and outflow of people in Wuhan.

*Jan. 25*: An 80 years old passenger disembarked from "Diamond Princess" cruise in Hong Kong, China. Then he had a fever.

*Feb. 1*: It is confirmed that the passenger who disembarked in Hong Kong has been infected with COVID-19.

*Feb. 3*: "Diamond Princess" cruise returned to Yokohama, Japan.

According to the actual voyage of "Diamond Princess" cruise, the **Basic scenarios** of the virus transmission model can be described as that the virus transmission on the cruise ship during the voyage when there is a virus carrier. The key parameters of basic scenarios can be set up as follows:

1) Due to the actual voyage of "Diamond Princess" cruise, the days of virus transmission ($N_{day}$) are set as 13 days in the model.

2) Adjust the protection ratio (percentage of persons who take individual protective measures) named $P_{protect}$. As a result, the infected situations of crowd with different values of $P_{protect}$ can be calculated.

3) Every scenario with one $P_{protect}$ is calculated for once.

Fig.3 shows the infection percentage of the crowd on the cruise with different values of $P_{protect}$ during the whole voyage. It can be illustrated that the virus infection pratio ($R_{crowd-vir}$) would arrive at its peak (~50%) without taking any protective measures. That is to say, because of the difference of individual immunity, the persons with weak resistance are all infected. Moreover, it can be seen that the virus infection ratio ($R_{crowd-vir}$) significantly decreases when the protective measures are taken. Thus, if the protective measures are completely taken (that is, every person takes the individual protective measures, $P_{protect}$=100%), the infection percentage can reduce to 2%.

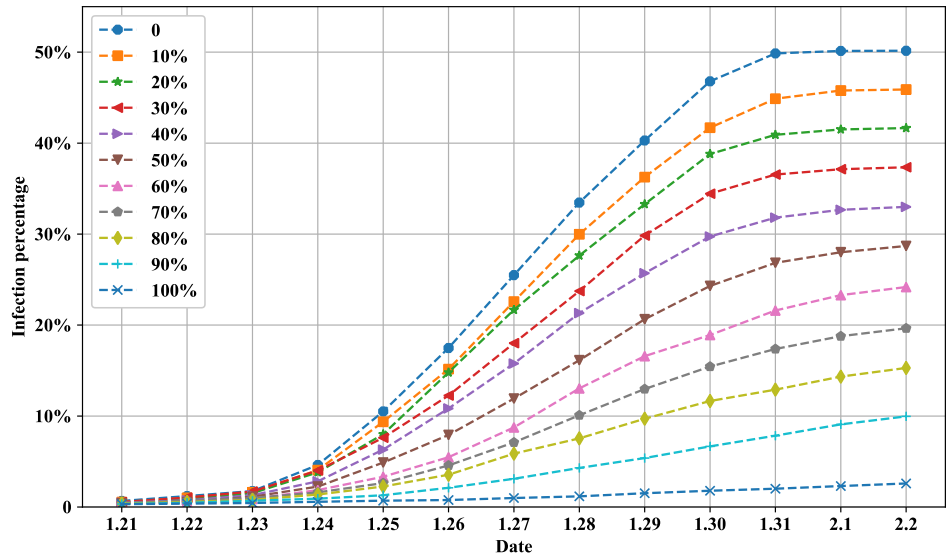

**Fig.3**. The infection percentage of crowd on cruise with different values of $P_{protect}$ during the whole voyage.

### 3.3 Self-protection Scenarios

Obviously, "Diamond Princess" cruise was not under the completely protective condition when the COVID-19 outbreaks. However, if "Diamond Princess" cruise don't take any protective measures, the virus infection ratio will be comparatively terrible. Will more cases be confirmed later?

Through careful investigation of some reports or news on the Internet, it can be found that some passengers have prepared some masks before boarding the cruise ship. Furthermore, people started to take some measures to strengthen self-protection gradually when they received the news that measures have been taken to limit inflow and outflow of people in Wuhan city on January 23, 2020. Therefore, the **Self-protection scenarios** are designed to simulate the process of virus infection under the strengthened self-protection condition. It should be a gradual process, that is, the $P_{protect}$ should increase day by day with the accumulation of epidemic information and the implementation of self-protection measures among the passengers and crew. The **Self-protection scenarios** are different from the **Basic scenarios**. For example:

1) The initial value of $P_{protect}$ is 10%;

2) The **Self-protection scenarios** with different $P_{protect}$ are described in Table.1. $P_{protect}$ increases by 10% per day from January 24, 2020 until it equals to a high threshold. As shown in Table. 1, six thresholds are employed in this model, which are 50%, 60%, 70%, 80%, 90%, 100% respectively. It should be emphasized that $P_{protect}$ doesn't increase any more when it equals to the threshold that set up in the scenarios.

3) The original virus carrier and the pedestrian who live in the same cabin disembark from the cruise ship on January 25, 2020 in this model. In other words, the values of $R_{per-vir}$ of these pedestrians are reset as 0 from January 26, 2020.

4) Every **Self-protection scenario** that described in Table.1 is calculated 5 times.

Table 1. Descriptions of the **Self-protection scenarios**

| ID | Description |
| --- | --- |
| Self - 50 | 1.21 ~ 1.23: $P_{protect}$=10%; 1.24 ~ 1.26: $P_{protect}$=20% ~ 40%; 1.27 ~ 2.2: $P_{protect}$=50% |
| Self - 60 | 1.21 ~ 1.23: $P_{protect}$=10%; 1.24 ~ 1.27: $P_{protect}$=20% ~ 50%; 1.28 ~ 2.2: $P_{protect}$=60% |
| Self - 70 | 1.21 ~ 1.23: $P_{protect}$=10%; 1.24 ~ 1.28: $P_{protect}$=20% ~ 60%; 1.29 ~ 2.2: $P_{protect}$=70% |
| Self - 80 | 1.21 ~ 1.23: $P_{protect}$=10%; 1.24 ~ 1.29: $P_{protect}$=20% ~ 70%; 1.30 ~ 2.2: $P_{protect}$=80% |
| Self - 90 | 1.21 ~ 1.23: $P_{protect}$=10%; 1.24 ~ 1.30: $P_{protect}$=20% ~ 80%; 1.31 ~ 2.2: $P_{protect}$=90% |
| Self - 100 | 1.21 ~ 1.23: $P_{protect}$=10%; 1.24 ~ 1.31: $P_{protect}$=20% ~ 90%; 2.1 ~ 2.2: $P_{protect}$=100% |

Fig.4 shows mean infection percentage curves of **Self-protection scenarios**. It can be demonstrated that the virus infection ratio ($R_{crowd-vir}$) would decrease effectively by improving the self-protection level spontaneously. It would be considered that the threshold of protection ratio ($P_{protect}$) of "Diamond Princess" cruise is between 60% and 80% even though it is impossible to obtain the data that described self-protection measures taken by people on the cruise. Thus, based on the infected situation of **Self-protection scenarios** on "Diamond Princess" cruise shown in Table.2, it can be concluded that the number of persons who are infected with COVID-19 will be between 850 and 1009 persons.

Table 2. The infected situation of **Self-protection scenarios** in "Diamond Princess" cruise

| ID | Infection rate | | Number of infections in the "Diamond Princess" |
|---|---|---|---|
| | Mean | Standard deviation | |
| Self - 50 | 0.3005 | 0.0034 | 1115.00 ± 17.23 |
| Self - 60 | 0.2720 | 0.0050 | 1009.47 ± 18.56 |
| Self - 70 | 0.2483 | 0.0087 | 921.52 ± 32.21 |
| Self - 80 | 0.2292 | 0.0099 | 850.73 ± 36.90 |
| Self - 90 | 0.2105 | 0.0051 | 781.13 ± 18.92 |
| Self - 100 | 0.1995 | 0.0121 | 740.17 ± 44.76 |

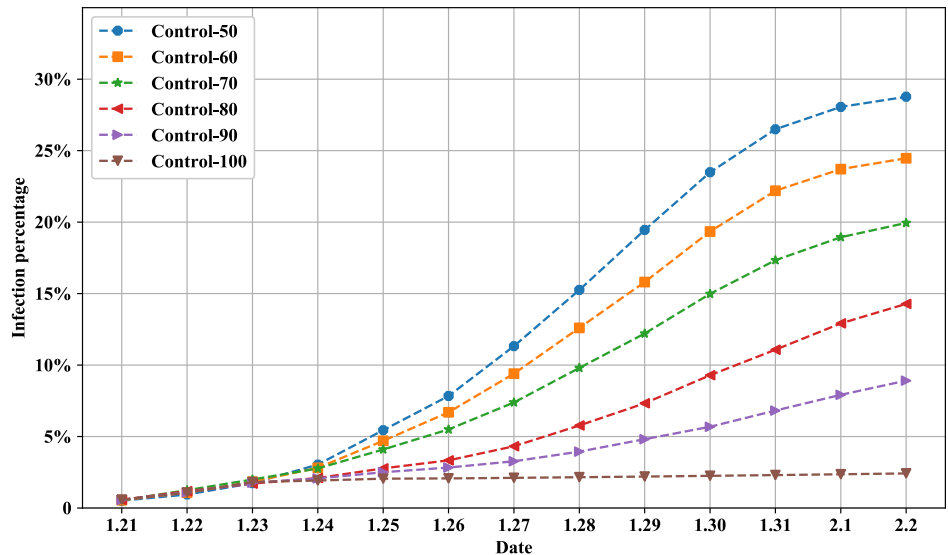

**Fig.4.** The mean infection percentage curves of self-protection scenarios.

### 3.4 Control scenarios

**Control scenarios** are set to simulate the condition that the managers of the cruise ship had begun to take recommended or mandatory measures to enhance protective level of people against the novel coronavirus since January 24. The mandatory measures should be work immediately. That's to say, the $P_{protect}$ would change to a certain high value immediately and keep it the same in the next few days with the mandatory measures are carried out. The differences between **Control scenarios** and **Basic scenarios** are:

1) The initial value of $P_{protect}$ is 10%;

2) The **Control scenarios** with different $P_{protect}$ are described in Table.3. $P_{protect}$ equals to a certain high value since January 24, 2020. As shown in Table. 3, six different values are adopted in this model, which are 50%, 60%, 70%, 80%, 90%, 100% respectively.

3) The original virus carrier and the pedestrian who live in the same cabin disembark from the cruise ship on January 25, 2020 in this model. In other words, the values of $R_{per-vir}$ of these pedestrians are reset as 0 from January 26, 2020.

4) Every **Control scenario** that described in Table.3 is calculated 5 times.

Table 3. Descriptions of the **Control Scenarios**

| ID | Description |
|---|---|
| Control - 50 | 1.21 ~ 1.23: $P_{protect}$=10%; 1.24 ~ 2.1: $P_{protect}$=50% |
| Control - 60 | 1.21 ~ 1.23: $P_{protect}$=10%; 1.24 ~ 2.1: $P_{protect}$=60% |
| Control - 70 | 1.21 ~ 1.23: $P_{protect}$=10%; 1.24 ~ 2.1: $P_{protect}$=70% |
| Control - 80 | 1.21 ~ 1.23: $P_{protect}$=10%; 1.24 ~ 2.1: $P_{protect}$=80% |
| Control - 90 | 1.21 ~ 1.23: $P_{protect}$=10%; 1.24 ~ 2.1: $P_{protect}$=90% |
| Control - 100 | 1.21 ~ 1.23: $P_{protect}$=10%; 1.24 ~ 2.1: $P_{protect}$=100% |

As shown in Fig.5, it can be found that the virus infection ratio would decrease effectively if the recommended or mandatory measures can be taken immediately during the alert phase of COVID-19 outbreaks. According to the infected situation of Control scenarios on "Diamond Princess" cruise described in Table. 4, provided that the final protection ratio ($P_{protect}$) equals to 80%, the final number of persons who are infected with COVID-19 will be less than the number that have been diagnosed as infected person on the cruise at present (691 cases as of February 24). In addition, it is expected that the virus infected number will be controlled in 90 persons if the value of $P_{protect}$ can reach 100%.

Table 4. The infected situation of control scenarios in "Diamond Princess" cruise

| ID | Infection rate | | Number of infections in the "Diamond Princess" |
|---|---|---|---|
| | Mean | Standard deviation | |
| Control - 50 | 0.2877 | 0.0026 | 1067.66 ± 9.78 |
| Control - 60 | 0.2446 | 0.0013 | 907.84 ± 4.77 |
| Control - 70 | 0.1994 | 0.0025 | 740.08 ± 9.42 |
| Control - 80 | 0.1428 | 0.0104 | 530.06 ± 38.42 |
| Control - 90 | 0.0891 | 0.0145 | 330.62 ± 53.74 |
| Control - 100 | 0.0242 | 0.0031 | 89.85 ± 11.32 |

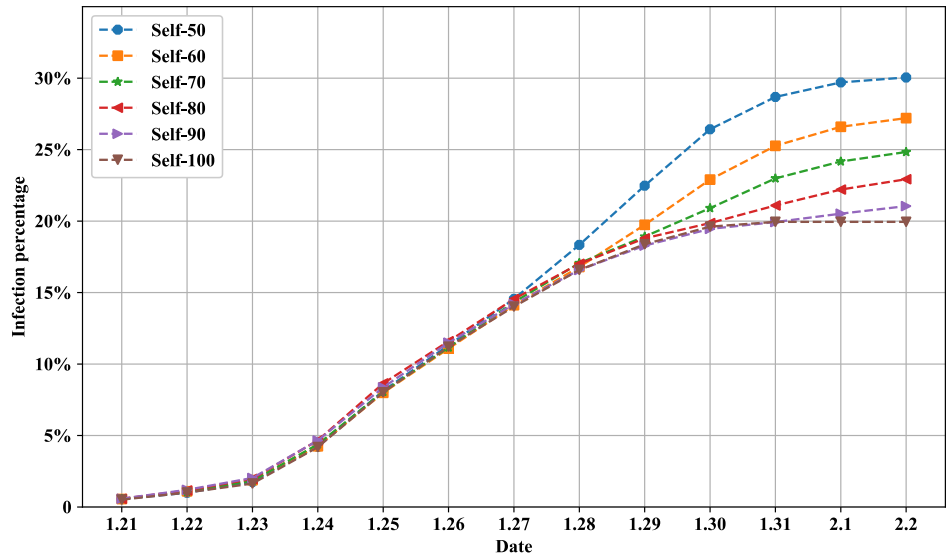

**Fig.5.** The mean infection percentage curves of control scenarios.

## 4. Summary

A virus transmission model based on the simulation of crowd flow was presented in this study, to simulate the transmission process of COVID-19 in a cruise ship, and then to judge how many infections there will be in the 3711 people on the "Diamond Princess" and analyze measures that could have prevented mass transmission.

In this article, three different types of scenarios of "Diamond Princess" cruise were designed to simulate the process of COVID-19 infection:

1) **Basic scenarios** simulated the virus transmission process among the crowd in the cruise ship with different protective ratios ($P_{protect}$). The results showed that if the individual protective measures were not taken in the confined cruise ship, the number of persons who would be infected with COVID-19 depends on the individual immunity during the 13-days voyage. That is, the persons with weak resistance are all infected.

2) **Self-protection scenarios** were designed to calculate the process of virus infection under the strengthened self-protection condition. The simulated situation would comparatively accord with the actual situation of "Diamond princess" cruise. According to the results obtained in the self-protection scenarios, the final number of persons who would be infected with COVID-19 will be between 850 and 1009 persons.

The actual number of persons diagnosed as virus-infected persons is less than the lower limit (850 persons) at present. However, it still should be emphasized that the patients on "Diamond Princess" cruise were mainly infected with COVID-19 during the voyage if the final infected number is close to the upper limit (1009 persons).

3) **Control scenarios** are set to simulate the effect of taking recommended or mandatory measures on virus transmission. It can be found that the final infected number would be controlled effectively if the recommended or mandatory measures can be taken immediately during the alert phase of COVID-19 outbreaks. Additionally, by comparing with the actual infected number, it can be indicated that the manager of the cruise ship didn't take appropriate mandatory measures.

**Acknowledgements**

This work was supported by the Key Research and Development Program (No. 2017YFC0803300), and Program of Shanghai Science and Technology Committee (18DZ1201500, 19QC1400900).

# Appendix

Animation 1 mainly shows the movement process of pedestrians going to the restaurant and going to the leisure area in the Self-50 scenario. The link of Animation 1 is: http://m.qpic.cn/psc?/V12nn6R53sqIuY/Nxhy4eqKGVy5xRHNFFSBrvXbOoPdZE9Lw7.fiB3*zejjdAjNbbvbuokVpvrtXY*hBNxKxbENCgi4ijDT*bpv0cSNW5rAM3wz9TKi96mN*xI!/b&bo=5QUqA.UFKgMCCS0!&rf=viewer_4

Animation 2 shows the movement process of pedestrians returning to their room. The link is: http://m.qpic.cn/psc?/V12nn6R53sqIuY/Nxhy4eqKGVy5xRHNFFSBrnBWQVgxjRm1yyTr2Ujl0aoVLjpZeC0qc.wthGwfaVNhcaEh3L.f*mMJ5Bt*xRq8KNTczLWiaULCeWOeGBlciW8!/b&bo=5QUqA.UFKgMCCS0!&rf=viewer_4

In these two animations, the symbol  represents the pedestrian without personal protective measure, and the symbol  represents the pedestrian with personal protective measure. The circles filled with different depths of color represent the pedestrians whose $R_{per-vir}$ are more than 10%, and the darker the color, the larger the value of $R_{per-vir}$.